\documentclass{emulateapj}
\usepackage{epstopdf}
\usepackage{amsmath, amsthm, amssymb} 
\usepackage{fancybox}
\usepackage{apjfonts}
\usepackage{graphicx} 
\usepackage{color} 
\usepackage{epsfig}
 
\newcommand{\be}{  \begin{eqnarray} }
\newcommand{\ee}{  \end{eqnarray} }

\def\spose#1{\hbox to 0pt{#1\hss}}
\def\lta{\mathrel{\spose{\lower 3pt\hbox{$\mathchar"218$}}
     \raise 2.0pt\hbox{$\mathchar"13C$}}}
\def\gta{\mathrel{\spose{\lower 3pt\hbox{$\mathchar"218$}}
     \raise 2.0pt\hbox{$\mathchar"13E$}}}

\begin{document}

\shorttitle{Super-Eccentric Jupiters}
\title{Super-Eccentric Migrating Jupiters }
\author{Aristotle Socrates\altaffilmark{1}, Boaz
  Katz\altaffilmark{1,2}, Subo Dong\altaffilmark{3} and Scott
  Tremaine}
\altaffiltext{1}{John N.\ Bahcall Fellow}
\altaffiltext{2}{Einstein Fellow}
\altaffiltext{3}{Sagan Fellow}

\affil{Institute for Advanced Study, Princeton, NJ 08540}
\begin{abstract}
\noindent
  An important class of formation theories for hot Jupiters involves
  the excitation of extreme orbital eccentricity ($e=0.99$ or even
  larger) followed by tidal dissipation at periastron passage that
  eventually circularizes the planetary orbit at a period less than 10
  days.  In a steady state, this mechanism requires the existence of a
  significant population of super-eccentric ($e>0.9$) migrating
  Jupiters with long orbital periods and periastron distances of only
  a few stellar radii.  For these super-eccentric planets, the periastron 
  is fixed due to conservation of orbital angular momentum and the energy dissipated per orbit is constant, implying 
that the rate of change in semi-major axis $a$ is $\dot{a}\propto a^{1/2}$
and consequently the number distribution satisfies
$d{\mathcal N}/d\log a\propto a^{1/2}$.  
If this formation process produces most hot
 Jupiters, {\it Kepler} should detect several super-eccentric
 migrating progenitors of hot Jupiters, allowing for a test of
high-eccentricity migration scenarios.
\end{abstract}

\section{Introduction}

The origin of gas-giant planets with orbital periods of only a few
days -- the hot Jupiters -- is not understood.  One hypothesis
involves the following sequence of events: (i) the planets form at a
few AU from their host stars, in approximately circular orbits; (ii)
some mechanism excites their orbits to extreme eccentricities
($1-e\lesssim 0.01$); (iii) tidal dissipation during successive
periastron passages removes enough orbital energy so that the planet
migrates a factor of $\sim 100$ in semi-major axis, finally settling
into a circular orbit close to the host star.  Possible excitation
mechanisms include Kozai--Lidov (KL) oscillations (Wu \& Murray 2003;
Fabrycky \& Tremaine 2007), planet-planet scattering (Rasio \& Ford
1996; Nagasawa et al.\ 2008), resonant capture during migration in
multi-planet systems (Yu \& Tremaine 2001), and weak resonant orbital
interactions (called ``secular chaos'' by Wu \& Lithwick 2011).  We
shall refer to these as high-eccentricity migration (HEM)
scenarios; they are of particular interest because they naturally
predict frequent misalignment of the stellar spin and planetary orbit,
consistent with recent observations of the Rossiter-McLaughlin effect
(Winn et al.\ 2010).

Most hot Jupiters have relatively small eccentricities: 60\% of the
known planets with orbital period $P<10\,\hbox{d}$ and $M\sin i >
0.25M_J$ (Jupiter masses) have eccentricities consistent with zero, and
90\% have eccentricity $e<0.1$. If the hot Jupiters are formed through
the HEM process, this result implies that the timescale for the decay
of the eccentricity from near unity to near zero is short compared to
the age of the Galaxy. Since the star-formation rate is approximately
constant over the age of the Galaxy, the distribution of migrating
Jupiters with moderate or high eccentricity should therefore be in an
approximate steady state.  Moreover, for these eccentricities the
energy dissipation per orbit is independent of eccentricity, since the
dissipation occurs only near periastron and in this region all
moderate and high-eccentricity orbits look like parabolae. Therefore,
we can predict the eccentricity and semi-major axis distribution in
HEM independent of the details of the dissipation process. We do this in
\S\ref{s: idea} and find that HEM requires the presence of a large
population of Jupiters with eccentricity $e\gtrsim 0.9$, which we call
``super-eccentric'' Jupiters.  Quantitative predictions and a 
strategy for detection using {\it Kepler} targets
are in \S\ref{s: conclude}.  A brief summary is given in 
\S\ref{s: summary}.

\section{Steady-State Distribution of Migrating Jupiters}
\label{s: idea}

\subsection{Basic assumptions}

\label{s:basic}

Migration from large to small semi-major axis requires that energy is
removed from the orbit.  In HEM, tidal friction is responsible for
converting orbital energy into heat which is then radiated away from
the system. In most cases, tidal friction in the planet removes energy
much faster than tidal friction in the star.  Since the
planet's spin angular momentum is negligible compared to its orbital
angular momentum, the orbital angular momentum per unit mass $J$ is
conserved during HEM (however, see discussion in
\S\ref{ss: Jorb}), i.e.,  
\be
 J^2=  G\left(M_{\star}+M_p\right)a\left(1-e^2\right)= {\rm cst.}
\label{e: J_const}
\ee  
where $a$, $M_p$, $M_\star$ and $G$ are the semi-major axis, planet mass,
host star mass and gravitational constant, respectively. Thus 
\be
a\,\left(1-e^2 \right)=q(1+e)\equiv a_{_{\rm F}}=
{\rm cst.}
\label{e: af_const}
\ee
where $q$ is the periastron distance and $a_{_{\rm F}}$ is the final
semi-major axis that the planet reaches 
when the eccentricity has decayed to zero. 

Let $X=e$ or $a$, and let ${\mathcal N}_J(X)\,dJ$ be the number of
migrating planets with eccentricity or semi-major axis less than $X$
and angular momentum in the interval $(J,J+dJ)$.  We assume that all
planets with eccentricity greater than some reference value $e_{\rm
  ref}$ are still migrating, and set 
${\mathcal N}_J(e_{\rm ref})=0$.  
We have argued that the distribution of migrating
planets is in steady state and that orbital angular momentum is
conserved during migration. Then the continuity equation requires 
\be
\dot{X}\,\frac{d {\mathcal N}_J}{dX}= {\mathcal S}_J
\label{e: simple_current}
\ee 
where ${\mathcal S}_JdJ$ is the current of migrating planets with
angular momentum in the interval $(J,J+dJ)$. This current is
determined by the properties of the source of highly eccentric
long-period gas giants, which is assumed to be far ($a\gg
1\,\hbox{AU}$) from the region of phase space under consideration. 

\subsection{Orbital evolution: approximate treatment at high eccentricity}
\label{sec:high_e}

We now describe an approximate analytic treatment of the orbital
evolution and steady-state distribution at high eccentricity.  For
high eccentricity the shape of the orbit near periastron and the
energy loss per periastron passage $\Delta E$ are both independent of $e$. Thus
the orbit-averaged energy loss rate is 
\be 
\dot{E}=\frac{\Delta
  E}{P}\propto \frac{1}{P} \propto a^{-3/2} \ee where $P$ is the
orbital period.  Since $E\propto 1/a$ \be
\Big|\frac{da}{dt}\Big|\propto a^{1/2}.
\label{e: adot}
\ee

In the region of $(e,a)$ space that contains a steady-state
distribution of migrating planets on high-eccentricity orbits
($q=a(1-e)\lesssim 10R_\odot$ for Sun-like host stars) the number of migrating Jupiters 
per unit semi-major axis is found with the help of 
equation (\ref{e: simple_current}),
\be
\frac{d{\mathcal N}_J}{da}=\frac{{\rm cst.}}{\dot{a}}\propto a^{-1/2}
\quad \mbox{or} \quad
\frac{d{\mathcal N}_J}{d\log a}\propto a^{1/2}.
\label{e:two}
\ee

\subsection{Orbital evolution: exact treatment}

\label{sec:exact}

In order to study orbital evolution at small or moderate eccentricity,
some understanding of tidal dissipation is required.  Unfortunately
there is no robust theory of tidal dissipation in gas-giant planets, due
both to the sparseness of observational calibration (only Jupiter and
Saturn) and to theoretical difficulties in studying such weak
dissipation (e.g., tidal $Q\sim 10^5$ for the Jupiter-Io system).
 
For illustration, we shall use the phenomenological approach of Hut
(1981), which follows Darwin in assuming that the tides lag their 
equilibrium value by a
constant time $\tau$.  By assuming pseudo-synchronous rotation (Hut's
eq.\ 45) we find that the
orbital evolution for a single planet is described by 
\be
\frac{de}{d\tilde{t}}=-\frac{1}{2}e(1-e^2)^{3/2}g(e)
\label{eq:eccev}
\ee
which is equivalent to
\begin{align}
\frac{d\tilde{a}}{d\tilde{t}} = &-\,\tilde{a}^{1/2}\,e^2g\left(e\right) 
\label{e: evolution}
\end{align}
where $\tilde{a}\equiv a/a_{_{\rm F}}=(1-e^2)^{-1}$ and $\tilde{t}\equiv
t/t_D$.  Here ${t_D= M_p\,a^8_{_{\rm F}}/\left(9\,k\,G\,M^2_{\star}\,R^5_p\,\tau \,
\right)}$ is a dissipation time\footnote{For planets of a fixed
density and a range of radii, the dissipation time scales as
$\tau\propto 1/(kR_p^2)$. Thus smaller planets have larger dissipation
times. Nevertheless, we expect the dissipation to occur mostly in the
planet rather than the star because the Love number $k$ is much
smaller in stars than planets.}, $k\simeq 0.5$ is the planet's
Love number, and $R_p$ is its radius. This result assumes $M_p\ll
M_\star$.
The function $g(e)$ is given by
\begin{align}
g(e)=&\frac{7+\frac{45}{2}e^2+56e^4+\frac{685}{32}e^6+\frac{255}{64}e^8+\frac{25}{256}e^{10}}{3(1+3e^2+\frac{3}{8}e^4)}
\nonumber \\
\simeq & \,2.33+6.12e^3
\end{align}
where the approximation in the final equation is accurate to better
than 0.5\% for all eccentricities between 0 and 1. 

\begin{figure}[t]
\epsscale{1.2}
\plotone{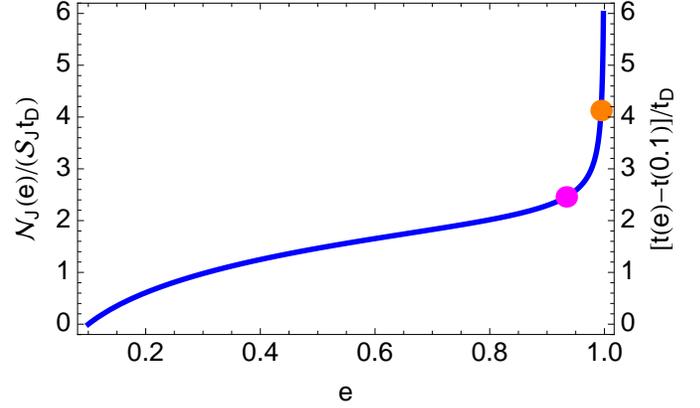}
\caption{Cumulative distribution of migrating planets ${\mathcal
    N}_J(e)$ with eccentricity $e>e_{\rm ref}=0.1$, from equation
  (\ref{e: cume_diss}). The planets migrate along a track of constant
  orbital angular momentum $J$. The vertical axis also represents the
  time required for the eccentricity to decay to $e_{\rm ref}$.  The
  magenta dot represents the current position of HD 80606b ($e=0.94$,
  $a=0.45$ AU) and the orange dot represents a hypothetical planet
  with semi-major axis $a=5$ AU flowing along the same
  angular-momentum track as HD 80606b.  The number of objects in the
  range $e=0.94-0.995$ is comparable to the number of objects in the
  range $e=0.2-0.94$. The plot is shown in normalized units so the
  curves are independent of $J$.}
\label{f: t_e}
\end{figure}

\begin{figure}[t]
\epsscale{1.2}
\plotone{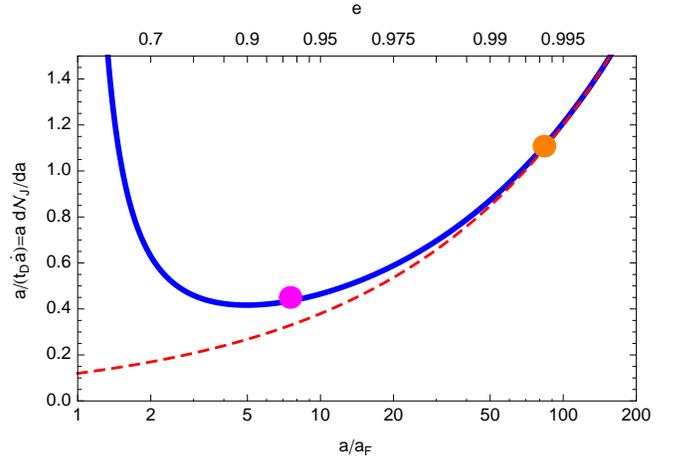}
\caption{The blue curve shows the distribution $d{\mathcal N}_J/d\log a$ of migrating Jupiters
  per unit interval in angular momentum $J$ from equation
  (\ref{e:one}) and the dashed red curve shows the high-eccentricity approximation 
  $d{\mathcal N}/d\log a\propto a^{1/2}$ (eq.\
\ref{e:two}). The high-eccentricity approximation is accurate to 
$\sim 20\%$  for $a/a_{_{\rm F}}\gtrsim 10$ or $e\gtrsim 0.95$.  The
magenta  and orange points have the same meaning as in Figure
\ref{f: t_e}.  The plot is shown in normalized units so the curves
  are independent of $J$.}
\label{f: dnda}
\end{figure}

Equation (\ref{e: simple_current}) then implies that the number of
planets per unit interval in angular momentum is given by
\be
\frac{d{\mathcal N}_J(e)}{de}=\frac{{\mathcal S}_J}{|de/dt|}, 
\label{e:one}
\ee
and 
\be
{\mathcal N}_J(e)=\int_{e_{\rm ref}}^e d{\mathcal
  N}_J(e)={\mathcal S}_J\left[t(e)-t(e_{\rm ref})\right]
\label{e: cume_diss}
\ee
where $t(e)$ is the time required to migrate from some initial
eccentricity near unity to $e$. The cumulative distribution ${\mathcal
  N}_J(e)$ for a single 
track in $J$, as determined
from equations (\ref{eq:eccev}) and 
(\ref{e: cume_diss}), is shown in Figure \ref{f: t_e}.

Figure \ref{f: dnda} displays the expected number of migrating
Jupiters per unit $\log a$, as obtained from equation (\ref{e:
  evolution}), as well as the high-eccentricity approximation
(\ref{e:two}).  For a fixed interval in orbital angular momentum, the
number of migrating Jupiters is an increasing function of $\log a$
above $e\simeq 0.9$.

\subsection{The approximation of constant orbital angular momentum}
\label{ss: Jorb}

In HEM scenarios, gas giants are assumed to be born on nearly circular orbits
and then acquire a large eccentricity after exchanging their angular momentum
with other planets or distant stellar companions, through close encounters in
the former case or Kozai--Lidov (KL) oscillations in the latter. 
Therefore, orbital angular momentum  is not a constant during the 
process of eccentricity excitation. Our analysis assumes that
eccentricity excitation takes place at large semi-major axes (say,
$a\gtrsim 5$--10 AU) and focuses on the region of $(a,e)$ space where
substantial orbital decay has already occurred but the eccentricity is
still moderate to large (say $a\lesssim 1$ AU and $e>0.2$).
It is not clear whether or not the approximation that migration takes
place at constant angular momentum is accurate 
for all semi-major axes $a\lesssim 1$ AU. 
In what follows, we assess the validity of the constant $J$ approximation
of \S\S \ref{sec:high_e} and \ref{sec:exact} in the presence of KL
oscillations, which are the most likely cause of changes in the
orbital angular momentum of the migrating planet. 

We performed many numerical integrations of the orbit-averaged 
restricted three body problem, including the effects of general 
relativity, tidal dissipation and tidal precession.  
Each simulation was initialized with a Jupiter-mass planet orbiting about a solar-mass star, placed in a
nearly circular orbit with semi-major axis $a\simeq 3-5$ AU. The
system also contained a solar-mass companion star, placed at distances of $30-1000$ AU
with inclination of $85^{\circ}\leq i\leq 90^{\circ}$ relative to the
planetary orbit.  Only the 
quadrupole term of the companion's potential was considered.

Typically, KL oscillations commenced at the start of the integration,
with large amplitude variations in
orbital angular momentum $J$.  Due to the strong dependence of tidal dissipation 
on periastron distance $q$, dissipation takes place almost entirely in 
the vicinity of $J_{\rm min}$, the minimum orbital angular momentum 
during a KL oscillation.  As a result, the value of $J_{\rm min}$ remains
roughly fixed during migration.  Precession due to general relativity acts
to decrease the amplitude of the oscillation in $J$ (e.g.,  Blaes et
al. 2002; Wu \& Murray 2003; Fabrycky \& Tremaine 2007).
Once the oscillation amplitude in $J$ is sufficiently small ($<10\%$ in $J$), such that the dissipation rate does not change considerably during each cycle, the mean value of $J$ remains constant and equal to the final 
orbital angular momentum $J_{\rm F}$.  Therefore, 
 the distribution of planets from then on can be computed by assuming 
 a constant $J=J_{\rm F}$.  At this stage of migration,  
 KL oscillations are considered to be "quenched."
Quantitatively, KL oscillations are quenched at a 
semi-major axis $a_{\rm Q}$ given by 
\be
a_{\rm Q}  \approx   1.8 {\rm AU}\,\left(\frac{a_{\rm F}}{0.05{\rm
      AU}}\right)^{-1/7}\left(\frac{\sin^2i_{\rm min}}{0.4} \right)^{-2/7}\left(\frac{M_{\star}}{M_{\odot}}\right)^{4/7}\nonumber\\
\times\left(\frac{M_{\rm per}}{M_{\odot}} \right)^{-2/7}
\left(\frac{a_{\rm per}}{1000\,{\rm AU}}\right)^{6/7}\left(\frac{1-e_{\rm
      per}^2}{1-0.5^2} \right)^{3/7} 
\label{e: quench}
\ee 
where $a_{\rm  per}$ and $e_{\rm per}$ are the semi-major axis and eccentricity of
the perturber and  $i_{\rm min}$ is the mutual inclination at the
phase of the KL oscillation when 
$J=J_{\rm min}$, while $M_{\rm per}$ is the
perturber mass.  For larger perturber mass or smaller semi-major axis,
KL oscillations are quenched closer to the host star.
In particular, nearby giant planets quench the oscillations at smaller
radii than distant companion stars; for example, a Jupiter-mass
perturber at $10$ AU has
$a_Q\approx 0.2$ AU.  The major difference between our
various integrations of KL oscillations with tidal dissipation was the
value of $a_Q$, due primarily to variations in the distance of the perturber.

During each integration we tracked the time that the planet spent in a
bin of width $\Delta J$ centered on the final angular momentum $J_{\rm
  F}$. We used this information to construct the eccentricity
and semi-major axis distributions that would be present in this
angular momentum bin in a
steady-state population of planets following this migration path. 
We found that even in the presence of KL oscillations 
the constant $J$ approximation described in \S\S \ref{sec:high_e} and
\ref{sec:exact} reproduced the distribution of planets to within a factor
of two or better, so long as $\Delta J$ was not more than about 20\%
of $J_{\rm F}$. 
That is, we found that the steady-state formulas derived 
in \S\ref{sec:exact} by assuming constant $J$ were still approximately valid, even though 
the orbital angular momentum experiences large amplitude 
oscillations (see example of HD 80606b in \S\ref{ss: example}).  

This surprising agreement results from the fact that tidal 
dissipation, and thus
migration, occurs mostly when $J\simeq J_{\rm min}$,
the minimum value of the angular momentum during a Kozai--Lidov cycle.  As long as
$J_{\rm min}$ is close to $J_{\rm F}$, the final value of 
orbital angular momentum after quenching,
then all of the migration takes place within the bin of width 
$\Delta J$. The time that the planets spend on Kozai--Lidov cycles
outside the bin is irrelevant, since they do not migrate there.

\subsection{An example}
\label{ss: example}

\begin{figure}
\epsscale{1.2}
\plotone{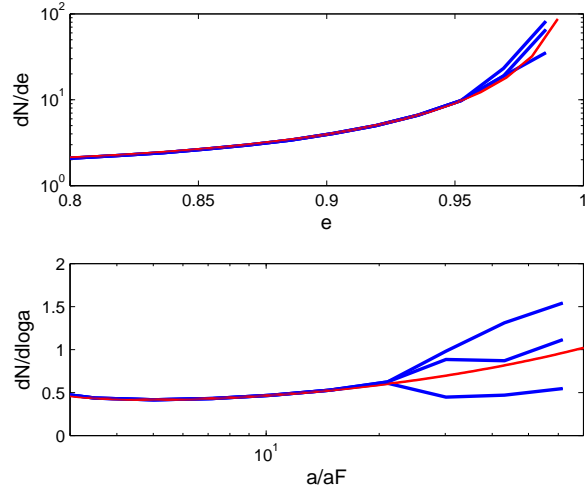}
\caption{ Comparison of the analytic theory (red) of \S\ref{sec:exact} with
  the migration of HD 80606b, as depicted in Figure 1 of Fabrycky \& Tremaine (2007). The blue lines show the density in eccentricity and log
semi-major axis of a
steady-state ensemble of planets that all follow the same trajectory
as HD 80606b. The planets only contribute to the density when their
angular momentum is small, in particular when $a\,(1-e^2)\le 0.14, 0.11$ and $0.08$ AU.
The normalization is chosen so that
all the curves match at $a/a_{\rm F}=5$.
Despite the presence of KL oscillations for $a/a_{_{\rm F}}\gtrsim
30$, the density is approximated by the analytic expressions
derived in \S\ref{sec:exact} to within a factor of two for all 
values of $(e, a)$.}
\label{f: Kozai}
\end{figure}

Consider the migration of the gas-giant planet HD 80606b, which
currently has $a=0.45$ AU and $e=0.93$, corresponding to $a_{_{\rm F}}=a(1-e^2)=0.06$ AU.
The migration track has been modeled by Wu \& Murray (2003) and
Fabrycky \& Tremaine (2007), who start with an initially nearly
circular orbit with $a = 5$ AU, similar to Jupiter. KL oscillations
are excited by a distant companion star HD 80607. For the first Gyr of
the evolution the eccentricity oscillates between $e_{\rm max}=0.993$
and $e_{\rm min}=0.04$--$0.25$. The amplitude of the KL oscillations
then gradually decays; the oscillations are quenched by 2.8 Gyr, when the
eccentricity is 0.97 and the semi-major axis is 2 AU -- in agreement 
with equation (\ref{e: quench}) --  and thereafter the eccentricity and semi-major
axis decay at constant angular momentum, reaching zero eccentricity
after 4 Gyr at a semi-major axis $a_{\rm F}=0.071$ AU. 

In Figures \ref{f: t_e} and \ref{f: dnda} the magenta and orange
points represent the current position of HD 80606b and its
hypothetical Jupiter-like ``progenitor,'' respectively.

Figure \ref{f: Kozai} shows in blue the density of planets in
eccentricity and semi-major axis that would result from a steady-state
ensemble of migrating planets with the same trajectory as HD
80606b. Three plots are shown, for angular-momentum cutoffs
$J_c^2=G(M_\star+M_p)a_c$ with $a_c=0.14,0.11$, and 0.08 AU (top to
bottom). For comparison, the red lines show the analytical predictions
of \S\ref{sec:exact}. In the latter stages of migration, after the KL
oscillations have been damped, the density matches the analytical
estimate extremely well---this is not surprising since the assumption
of evolution at constant angular momentum is satisfied to high
accuracy. At larger eccentricities and semi-major axes, when KL
oscillations are present, the blue curves are displaced from
extrapolation of these theoretical predictions by up to a factor of
two or so, but their shapes remain similar as expected from the
arguments of the preceding subsection. 

All three of Figures \ref{f: t_e}, \ref{f: dnda} and \ref{f: Kozai}
show that in a steady state, an unbiased sample of exoplanets
containing one HD 80606b should contain more than one migrating planet
with a similar periastron distance and even larger semi-major axis and
eccentricity.

\begin{figure}
\epsscale{1.2}
\plotone{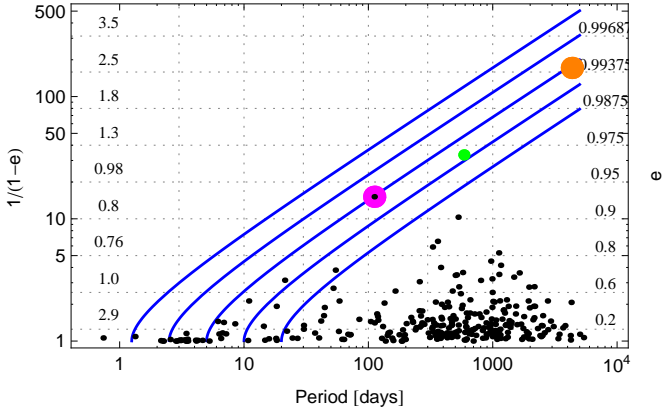}
\caption{ 
In solid blue are lines
of constant angular momentum $J$ that correspond to 
$P_{_{\rm F}}=P(1-e^2)^{3/2}=1.25,\, 2.5,\, 5.0,\, 10.0,\, {\rm and} 
 \,20.0$ days.  Horizontal dotted lines 
are fixed values of $1/1-e$, separated by factors of two.  The 
corresponding values of $e$ (0.2, 0.6, etc.,) are on the right hand side
of the figure.  The relative
number of planets expected in each interval along a track in 
constant $J$ is given by the numbers on the left (2.9, 1.0, etc.,).
Tidal dissipation drives the planets to flow from the upper right corner 
to the lower left corner, along the lines of constant $J$.
Black dots are current RV observations  with 
$M\sin i> 0.25$ taken from exoplanet.org  with the
exception of HD 20782b (green dot) for which the updated
$e=0.97$ (O'Toole et al.\ 2009) is used. 
The magenta dot highlights the current position of 
HD 80606b and the orange dot represents a 
hypothetical planet
at $5$ AU flowing along the same track in $J$ as HD 80606b. }
\label{f: flow}
\end{figure}

\section{Observations, Predictions, 
and Discussion}
\label{s: conclude}

\subsection{Current observations}

We compile a list of all known exoplanets with $M\sin i>0.25M_J$, of
which the radial-velocity planets are displayed in Figure \ref{f:
flow}. The planetary parameters are taken from {\tt exoplanets.org}
with the exception of HD 20782b, the green dot, whose eccentricity was
recently revised to $e=0.97$ (O'Toole et al.\ 2009). For each planet we
compute $P_{_{\rm F}}=P(1-e^2)^{3/2}$, the final orbital period that a
planet would reach if its eccentricity decayed to zero at constant
angular momentum.  For reference a planet with Jupiter's
orbital period and $P_{_{\rm F}}=10\,\mbox{d}$ would have $e=0.991$.

The blue lines in Figure \ref{f: flow} are lines of constant 
orbital angular momentum, along which planets flow from 
long to short orbital periods.  The relative
number of planets expected in each interval along a track in 
constant $J$ is given by the numbers on the left (2.9, 1.0, etc.,).
If for example, 
one gas giant planet is found migrating in the $P_{\rm F}$ = 5-10 d bin within 
an eccentricity range $0.9<e<0.95$ (such as HD 80606b), then 
there should be $\gtrsim 1$ planet migrating in this 5-10 d bin within an
eccentricity range $0.95<e<0.975$ as well.

Among the known gas-giant exoplanets there is a significant excess
population having eccentricity consistent with zero and
$P=P_{_{\rm F}}<10\,\hbox{d}$, corresponding to $a<0.09\,\mbox{AU}$ for a
solar-mass host star. In the HEM scenario, these are planets that were
formed at several AU, excited to high eccentricity, migrated due to
tidal friction, and have now completed the migration process. There is
no such excess for larger periods; in HEM models this implies that tidal dissipation
is unimportant for planets with $P_{_{\rm F}}>10\,\mbox{d}$ and we
discard these from our sample.

{}From the remaining sample we calculate the number of ``moderately
eccentric'' planets, which we define to be those with $0.2<e<0.6$, and
the number of ``super-eccentric'' planets ($e>0.9$).  The first two
lines of Table \ref{t: predictions} summarize the number of gas-giant
planets with moderate eccentricity, as detected by radial-velocity
(RV) surveys and transit photometry with spectroscopic follow-up
(``Transit+RV'')\footnote{The planets in the ``Transit+RV'' line in
  Table 1 are obtained by using the following search string in {\tt
    exoplanets.org} (Wright et al.\ 2011): 
    \be {\rm MSINI\,[mjupiter]\, > 0.25\,\, and\,\,
    PER\,[day] * (1 - ECC^2)^{3 / 2} > 3} \nonumber\\ {\rm and\,\,
    PER\,[day] * (1 - ECC^2)^{3 / 2} < 10\,\, and \,\,DISCMETH ==
    ``TRANSIT"}\nonumber
    \ee They mostly consist of planets discovered in
  ground-based transit surveys (over 70\%) as as well as a handful of
  objects discovered by the {\it COROT} and {\it Kepler} space-based
  telescopes and by RV surveys. All objects in this category have eccentricities that
  have been determined by spectroscopy.  The vast majority of {\it
    Kepler} planets have no spectroscopic follow-up and hence are not
  included in this line.}.  Both the RV and Transit+RV categories
yield a fraction of moderate-eccentricity planets that is roughly
$1/2$ in the 5--10 d bin and much smaller, $\lesssim 1/15$, in the 3--5 d
bin.  The sharp decline for smaller values of $P_{_{\rm F}}$ is
consistent with the expectation that tidal dissipation is stronger for
orbits with smaller periastron, so that in a steady state the fraction
of planets in the migration pipeline is smaller.

In HEM models both the moderately eccentric and super-eccentric
planets are in steady-state migration and therefore the population
ratio in these groups can be calculated using the models of
\S\ref{sec:exact}. Thus we can predict the number of super-eccentric
planets that should have been found in RV surveys; this prediction is
shown in boldface in the third line along with the number of
super-eccentric planets actually found so far in these surveys.
Furthermore, the fractions of moderately
eccentric RV and Transit+RV planets can be used with our models to
predict the number of super-eccentric planets in the {\it Kepler}
sample (Borucki et al.\ 2011); this prediction is shown in boldface in
the last line of the Table. 

Before discussing these predictions we address selection effects.  RV
surveys may be biased against the detection of super-eccentric planets
for at least two distinct reasons.  First, sparse observations of
high-eccentricity orbits are likely to miss the strong reflex velocity
signal near periastron, leading to non-detection of planets that would
be detected at the same semi-major axis and smaller eccentricity, or
to an underestimate of the eccentricity if the planet is detected
(Cumming 2004, O'Toole et al.\ 2009).  This bias only sets in for $e\gtrsim 0.6$ so
the fraction of moderately eccentric planets detected in RV surveys is
much more reliable than the fraction of super-eccentric
planets. Second, we are mostly concerned with avoiding biases against
detecting highly eccentric planets at a given angular momentum (i.e.,
along a given migration track), rather than at a given semi-major
axis. Here there is an additional bias, since high-eccentricity orbits
have longer periods and hence a periodic signal is harder to detect
and characterize in a given time baseline.

Despite these poorly understood selection effects, the predicted and
observed numbers of super-eccentric planets in RV surveys as shown in
Table 1 are consistent. However, the numbers are too small to test the
validity of HEM scenarios.
 
 \begin{deluxetable}{ccc}
\tabletypesize{\scriptsize}
\tablecaption{Expected number of super-eccentric planets 
\tablenotemark{a}
\label{t: predictions}}
\tablewidth{0pt}

\tablehead{
\colhead{ ${P_{_{\rm F}}}=  P(1-e^2)^{3/2}$} & \colhead{3--5 days} 
& \colhead{5--10 days }  }
\startdata
RV (moderate/total) & 0/13 & 4/9 
\\
Transit+RV (moderate/total) & 3/46 & 3/8 
\\
RV (super-eccentric, {\bf theory}/observed)& {\bf 0} vs. 0 & {\bf 2}--{\bf 3 } vs. 2
\\
{\it Kepler}   (super-eccentric, {\bf theory}) & {\bf 2} &{\bf 3}-- {\bf 5} 
\enddata 
\tablenotetext{a}{``Moderate'' denotes the eccentricity range
  $0.2<e<0.6$, ``super-eccentric" denotes $e>0.9$, and ``total'' is
  $0\le e<0.6$. Numbers in boldface are predictions obtained from the
  number of moderate-eccentricity planets in the RV and Transit+RV categories by assuming that all giant
  planets at small periods are formed by high-eccentricity migration
  (HEM) and applying the model of \S\ref{sec:exact}. The period
  intervals (e.g., 3--5 days) refer to the final period $P_{_{\rm
      F}}=P(1-e^2)^{3/2}$, which is the period after HEM is complete
  and the orbit is circularized, assuming constant orbital angular
  momentum.  Only planets with $M\sin i > 0.25\, M_J$ are included in
  the statistics. The predictions are for super-eccentric planets with
  orbital period $P<2$ yr only. The results are based on queries to
  the {\tt exoplanets.org} database in  September 2011. The
  predictions do not account for eccentricity-dependent selection effects.}
\end{deluxetable}

For transit surveys the selection effects can be divided into
geometric effects, which depend on the orientation of the observer
relative to the star (i.e., whether or not a planet transits the star), and
survey effects, which depend on properties of the survey (time
baseline, photometric accuracy, etc.). With adequate baseline and signal-to-noise ratio, the
most important selection effect is geometrical: the probability that the planet will transit is given by
\be
{\mathcal P}=\left\langle R_\star/r\right\rangle_\phi
=\frac{R_{\star}}{a\,\left(1-e^2\right)}
=\frac{R_{\star}}{a_{_{\rm F}}} \propto\frac{R_{\star}}{J^2}.
\ee
where $R_{\star}$ and $r$ are the stellar radius and heliocentric 
distance of the planet during transit and $\langle\rangle_{\phi}$
is an average over the azimuth of the sightline, 
which is equal to an average over the true anomaly.
Therefore, {\it on a migration track of constant angular momentum, 
the geometric selection effects are independent of eccentricity.}  
That is, a hot Jupiter progenitor with say, $a=1$ AU and $e=0.975$
has the same detection probability as a circularized hot Jupiter
with  $a=0.05$ and $e=0$.  Survey selection effects, in contrast, are biased against
high-eccentricity orbits because the period is longer so there are
fewer transits in a given period, and because the transits are shorter
so the signal/noise ratio is smaller. However, these selection effects
can be calculated and corrected for using the methods outlined in Borucki et
al. (2011), and should be relatively small since our sample is
restricted to giant planets, which are relatively easy to detect.

\subsection{Predictions  for {\it Kepler}}

{\it Kepler} observations yield the planetary radius $R$ and orbital
period $P$. To compare these results to our models we assume that our
mass limit, $0.25M_J$, corresponds to $R=8R_\oplus$ and that the
planet population within a range of period is roughly the same as the
population within the same range of $P_{_{\rm F}}=P(1-e^2)^{3/2}$
since most planets have small eccentricities. The most recent {\it
  Kepler} catalog (Borucki et al.\ 2011) contains 30 planets with $R\ge
8R_\oplus$ and 3 d$\,\le P\le 5\,$d, and 16 in the same radius range
with 5 d$\,\le P\le 10\,$d.\footnote{Note that the ratio of planets in
these two period bins, $16/30=0.5$, is larger than the corresponding
ratio for ground-based surveys, $8/46=0.2$ (the numbers are the same
whether we use $P$ or $P_{\rm F}$). This result suggests that {\it Kepler} 
has less selection bias against long-period gas giants
than ground-based surveys, which favors the detection of
super-eccentric migrating planets.}
 
   These can be combined with
the results from ground-based surveys (line 2 of Table 1) to predict
the number of moderate-eccentricity planets in each period range, and
these numbers are combined with the steady-state HEM models in
\S\ref{sec:exact} to predict the numbers of super-eccentric planets in
the {\it Kepler} catalog (line 4 of Table 1).  These predictions
should be underestimates since {\it Kepler} will detect planets with
longer periods as the mission progresses (the automated
pipeline in Borucki et al.\ 2011 only finds objects with $P < 93$ d).

These results imply that {\it Kepler} should detect several
super-eccentric ($e>0.9$) giant planets ($R>8R_\oplus$) with orbital
period <2 yr. If an extended {\it Kepler} mission permits detections
of planets with longer periods the predicted number is higher.  A
significant fraction of these could have 
 $e> 0.94$ i.e., more eccentric than HD 80606b, the current
confirmed record-holder.

A typical member of this population, with $a_{_{\rm F}}= 0.1$ AU and
$M>0.25M_J$, produces a stellar reflex velocity $> 50\,\mbox{m
  s}^{-1}$ near periastron.  For objects on highly eccentric orbits
with random orientations, most transits occur near periastron, where
the reflex velocity is close to the periastron value---quantitatively,
over half of all transits occur when the reflex velocity is within
10\% of the periastron velocity. Thus relatively few low-exposure RV
measurements near the transit epoch should be sufficient to detect and
measure a large eccentricity.  We suggest that all of the {\it Kepler}
gas-giant planetary candidates with periods above $\gtrsim 20$ d
(Borucki et al.\ 2011 list $34$ objects with $R>8R_\oplus$ and periods
between 20 d and 93 d) be followed spectroscopically near transit
(with one or two additional measurements at other phases to determine
the systemic velocity).

\section{Summary and Discussion}\label{s: summary}

The main result of this paper is that if hot Jupiters are formed by
high-eccentricity migration (HEM), then there must be a steady-state
current or flow of gas-giant planets migrating from large to small
orbital periods.  Since tidal dissipation is required for HEM and is
only effective out to distances of a few stellar radii in typical
exoplanet systems, the current must consist of planets that either
have periastrons of a few stellar radii or undergo Kozai--Lidov
oscillations or other dynamical processes that regularly bring their
periastrons to these small values. Moreover, because energy loss from
tidal dissipation only occurs near periastron, the rate of energy loss
on high-eccentricity orbits varies inversely with the orbital period;
thus for every migrating planet on a moderate-eccentricity orbit there
should be many super-eccentric planets ($e>0.9$). We have computed the
expected eccentricity and semi-major axis distribution of the
steady-state current of migrating planets using Hut's (1981) model of
tidal dissipation and assuming pseudo-synchronous planetary spin. Our
results indicate that several super-eccentric gas-giant planets should
be present in the {\it Kepler} exoplanet catalog. These can be
discovered, if present, by a program of radial-velocity measurements
on the Kepler planets with the largest diameters and the longest
periods. 

The absence of a significant number of super-eccentric migrating
Jupiters in this sample would imply either that HEM is not an
ingredient of the formation process for most hot Jupiters, or that our
migration model is oversimplified. In particular, we assume that
migration occurs at constant orbital angular momentum but argue that
our results should be approximately correct even in the presence of
Kozai--Lidov oscillations or other processes.

The simple HEM model described here, whose central components are the
steady-state approximation and the assumption that migration occurs at
constant angular momentum, provides a preliminary framework for the
exploration of the dynamics of HEM. A thorough exploration of this dynamics should
establish whether our simplified model is accurate and enable a
definitive observational test of whether hot Jupiters form through
HEM.

\acknowledgements{We thank Dan Fabrycky and Andy Gould for useful discussions. This
  research was supported in part by NASA grant NNX08AH83G.  
  BK is supported by NASA through the Einstein Postdoctoral Fellowship awarded by Chandra X-ray Center, which is operated by the Smithsonian Astrophysical Observatory
  for NASA under contract NAS8-03060.  
  Work by SD was performed under contract with the California Institute
of Technology (Caltech) funded by NASA through the Sagan Fellowship
Program.  This research has made use of the Exoplanet Orbit Database
and the Exoplanet Data Explorer at {\tt exoplanets.org}.}


\begin{thebibliography}{}


\bibitem[Blaes et al.(2002)]{2002ApJ...578..775B} Blaes, O., Lee, M.~H., 
\& Socrates, A.\ 2002, \apj, 578, 775 

\bibitem[Borucki et al.(2011)]{2011ApJ...736...19B} Borucki, W.~J., et al.\ 
2011, \apj, 736, 19 

\bibitem[Cumming(2004)]{2004MNRAS.354.1165C} Cumming, A.\ 2004, \mnras, 354, 1165 

\bibitem[Fabrycky \& Tremaine(2007)]{2007ApJ...669.1298F} Fabrycky,
  D., \& Tremaine, S.\ 2007, \apj, 669, 1298

\bibitem[Hut(1981)]{1981A&A....99..126H} Hut, P.\ 1981, \aap, 99, 126 

\bibitem[Nagasawa et al.(2008)]{2008ApJ...678..498N} Nagasawa, M., Ida, S., 
\& Bessho, T.\ 2008, \apj, 678, 498

\bibitem[O'Toole et al.(2009)]{2009MNRAS.392..641O} O'Toole, S.~J., Tinney, 
C.~G., Jones, H.~R.~A., Butler, R.~P., Marcy, G.~W., Carter, B., 
\& Bailey, J.\ 2009, \mnras, 392, 641 

\bibitem[Rasio  \& Ford(1996)]{1996Sci...274..954R} Rasio, F.~A., \& Ford, E.~B.\ 1996, Science, 274, 954 

\bibitem[Winn et al.(2010)]{2010ApJ...718L.145W} Winn, J.~N., Fabrycky, D., 
Albrecht, S., \& Johnson, J.~A.\ 2010, \apjl, 718, L145

\bibitem[Wright et al.(2011)]{2011PASP..123..412W} Wright, J.~T., et al.\ 2011, \pasp, 123, 412 

\bibitem[Wu \& Lithwick(2011)]{2011ApJ...735..109W} Wu, Y., \&
  Lithwick, Y.\ 2011, \apj, 735, 109

\bibitem[Wu \& Murray(2003)]{2003ApJ...589..605W} Wu, Y., \& Murray,
  N.\ 2003, \apj, 589, 605

\bibitem[Yu \& Tremaine(2001)]{2001AJ....121.1736Y} Yu, Q., \&
  Tremaine, S.\ 2001, \aj, 121, 1736

\end{thebibliography}
\end{document}